\documentstyle[aps,multicol,epsf]{revtex}

\draft
\begin{document}
\title{Statistics of Largest Loops in a Random Walk}
\author{Deniz Erta\c s{\dag} and Yacov Kantor{\ddag}}
\address{
{\dag}Department of Physics, Harvard University, Cambridge, MA 02138, 
U.S.A. \\
{\ddag}School of Physics and Astronomy, Tel Aviv University, 
Tel Aviv 69 978, Israel
}
\date{\today}
\maketitle
\begin{abstract}
We report further findings on the size distribution of
the largest neutral segments in a sequence of $N$ randomly charged 
monomers [D. Erta\c s and Y. Kantor, Phys. Rev. E{\bf 53}, 846 (1996)]. 
Upon mapping to one--dimensional random walks (RWs), this corresponds 
to finding the probability distribution for the size $L$ of the
largest segment that returns to its starting position in an $N$--step RW.
We primarily focus on the large $N$, $\ell= L/N \ll 1$ limit, which exhibits
an essential singularity. We establish analytical upper and lower 
bounds on the probability distribution, and  numerically probe the 
distribution down to $\ell \approx 0.04$ (corresponding to probabilities
as low as $10^{-15}$) using a recursive Monte Carlo algorithm.
We also investigate the possibility of singularities 
at $\ell=1/k$ for integer $k$. 
\end{abstract}

\begin{multicols}{2}

\section{Introduction}
It has recently been shown that ground state conformations of
polyampholytes, a particular type of heteropolymers built with
a random mixture of positively and negatively charged groups
along their backbone, are extremely sensitive to the their
total (excess) charge $Q$. A detailed study of the 
$Q$--dependence of the radius of gyration $R_g$\cite{KK,KKenum}
determined that a reasonable compromise
between stretching (which minimizes the electrostatic energy) 
and remaining compact (which gains in condensation energy) 
is for the polyampholyte to form a {\it necklace} of weakly 
charged blobs connected with highly charged ``necks'', by taking  
advantage of the charge fluctuations along the chain. The 
results of Monte Carlo\cite{KK} and exact enumeration\cite{KKenum} 
studies qualitatively support such a picture. 

\begin{figure}
\narrowtext
\centerline{\epsfxsize=2.9truein\epsffile{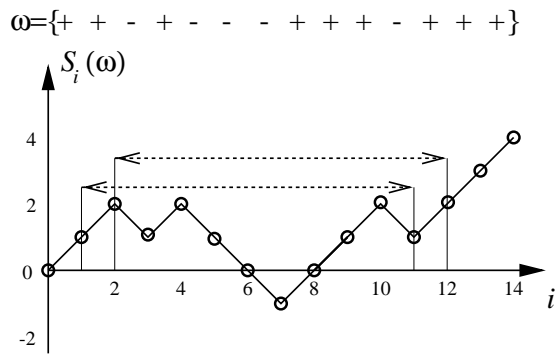}}
\medskip
\caption{Example of a sequence $\omega$ with $N=14$ charges, 
and the corresponding walk
depicted by $S_i(\omega)$. In this case, the longest loops 
have lengths $L=10$ (dotted lines).}
\label{randomwalk}
\end{figure}

While the exact treatment of electrostatic interactions
is not possible, we can pose a simplified problem which, 
we hope, captures some essential features of this necklace
model. For example, we may ask what the typical
size of the largest neutral (or weakly charged) segment  
in a random sequence of $N$ charges will be. 
In order to answer this question, we investigated
the size distribution
of the largest neutral segments in 
polyampholytes with $N$ monomers ($N-$mers).  
This problem can be mapped to a one-dimensional
random walk (RW): the sequence of charges $\{q_i\}$
($i=1,\dots,N;\  q_i=\pm1$) corresponds to an $N$--step walk
$\omega\equiv\{q_1,\cdots,q_N\}$
with the same sequence of unit steps in the positive or negative 
directions along an axis, where the probability of going up or 
down is equal to $1/2$ at each step. 
Fig.~\ref{randomwalk} depicts an example 
of such a sequence and the corresponding path, where 
 $S_i(\omega)=\sum_{j=1}^iq_j$ is the position
of the path at index $i$. 
($S_0(\omega)\equiv0$.)
A segment of $L$ monomers with zero total charge  
thus corresponds to an $L$--step loop inside the RW. 
In this paper, we  further investigate properties of the 
probability $P_N(L)$ that the {\it largest} loop in an $N$--step
RW has length $L$, or, equivalently, the probability
$Z_N(L)=\sum_{L'=0}^{L-1} P_N(L')$ that all loops in an 
$N$--step RW are shorter than $L$. Earlier results about
a generalized version of this and other related problems 
can be found in Refs.~\cite{KEloop,EKlongloop}. 
In the continuum ($N\to\infty$) limit, it is more 
convenient to work with the {\it probability density}
\begin{equation} 
p(\ell)\equiv{N\over 2}\left[P_N(L)
+P_N(L+1)\right]
\end{equation} 
and
\begin{equation}
z(\ell)=\int_0^\ell d\ell' p(\ell'),
\end{equation}
where $\ell=L/N$ is the appropriate scaling variable for this problem.

There is an apparent simplicity of the formulation of the
problem, i.e. it is similar (and related) to the classical
RW problems\cite{rchandra},  such as the problem
of first passage times or the problem of last return to the
starting point, for which probability
distributions can be computed exactly by using the method
of reflections\cite{mathnote}, and obey the same scaling in the
continuum limit. However, the search for the
{\it longest} loop of the RW, among all possible starting points, 
creates a more complicated problem. In its essence, the problem 
is more related to the  
statistics of self--avoiding, rather than regular, random walks.
This relation becomes more transparent in the $\ell\to1$ and 
$\ell\to 0$ limits. The former limit had been extensively studied
in Ref.~\cite{EKlongloop}, and the latter will be discussed in
Sec.\ref{numerical}. The ``self--interacting 
nature" of the problem can be seen even more clearly in its generalizations 
to arbitrary space dimension $d$, where many  
analogies between this problem and the self--avoiding walks exist.

Our earlier investigations revealed remarkable properties
of the probability density $p(\ell)$: It diverges as 
$p(\ell)\sim1/\sqrt{1-\ell}$ for $\ell\to 1$, and has a  
discontinuous derivative at $\ell=1/2$.
Furthermore, it has an essential singularity at $\ell=0$ of the 
form $p(\ell)\sim \exp(-B/\ell)$. An analytical solution in 
this limit still remains elusive. We had not been able to 
determine $p(\ell)$ even numerically below $\ell\approx 0.15$ due to
the very small probabilities involved near $\ell=0$, severely
limiting a straightforward Monte Carlo approach. Because of these
difficulties, the existence and precise form of this singularity
(including possible power law prefactors etc.) was not well 
established.
Since the publication of that work, we have developed an improved
Monte Carlo algorithm that is capable of probing 
significantly smaller values of $\ell$ numerically. Combined
with strict analytical bounds on $z(\ell)$, the results strongly
favor the existence of this singularity, and the proper form
of the $\ell\to 0$ limit 
can be determined with high precision. In this paper, we report 
the results of these complementary findings.

It should be noted that similar behavior is exhibited
by extremal properties of a number of random processes,
such as a one-dimensional random cutting process\cite{Derrida} 
(which can be generalized to higher dimensions\cite{Frach}) and 
return times in a random walk\cite{Frach}. These models exhibit 
singularities at $\ell=1/k$, which become progressively weaker as 
the integer $k$ is increased, leading to an essential 
singularity at $\ell=0$. 
Although it was claimed that our 
problem falls into the same category and therefore should exhibit 
singularities at $\ell=1/2,1/3,1/4,\cdots$\cite{Frach}, we believe
that it differs from these models in a way that undermines
the reasoning for this claim, as we shall discuss in Sec.~\ref{related}. 
In particular, we have numerically verified that the suggested 
singularity at $\ell=1/3$ does not exist, unless it has a very 
small prefactor.

The rest of the paper is organized as follows: First, we establish 
upper and lower bounds on $z(\ell)$. We then describe an efficient
Monte Carlo algorithm that enables us to determine $z(\ell)$ down
to very small values, and present results from its implementation. 
Finally, we discuss the possible relevance of other random
models with similar characteristic properties.

\section{Upper and lower bounds}

In this section, we establish rigorous upper and lower bounds on
the probability distribution $z(\ell)$, both of which have the
same functional form. The existence of these bounds significantly
restrict possible asymptotic forms of $z(\ell)$ in the $\ell\to 0 $ limit.

The main strategy is the similar for establishing both upper and lower bounds. 
Walks whose largest loops are much smaller than their overall length 
are typically very biased in one direction, and sections of the walk
that are separated by more than the largest loop size are very weakly
correlated. For a given (small) value of $\ell$, let us divide each walk 
into roughly $1/\ell$ segments of similar size. There are 
{\it necessary} conditions that each segment must satisfy independently
for the overall walk to contribute to $z(\ell)$. If the probability
for a random segment to satisfy these conditions is $p_n$, then 
$z(\ell)>p_n^{1/\ell}$. Similarly, each segment can be designed to
satisfy certain conditions that are {\it sufficient} to ensure that the
overall walk contributes to $z(\ell)$. If the corresponding probability
for these conditions is $p_s$, then $z(\ell)<p_s^{1/\ell}$. The rest of
this section is devoted to establishing a set necessary and sufficient 
conditions and calculating the corresponding probabilities.

Let us first investigate necessary conditions.    
Let $\omega$ be an $N$--step walk whose largest loop is less 
than $L$--steps long, and 
has $S_N(\omega)>0$. We shall focus on the cases
where $m=N/L$ is an integer for now. 
Let us split $\omega$ into $m$ mutually 
exclusive segments $\{\omega_1,\cdots,\omega_m\}$ of length $L$ 
where $\omega_i=\{q_{(i-1)L+1},\cdots,q_{iL}\}$. 
It is easy to see that $\omega$ satisfies the inequalities
\begin{equation}
S_{iL}(\omega)>S_{(i-1)L}(\omega),\; 0 < i \leq m,
\end{equation}
or, equivalently,
\begin{equation}
S_{L}(\omega_i)>0,\; 0 <  i \leq m,
\end{equation}
i.e. each of the $m$ segments need to have a positive displacement. 
The probability for this is just $p_n=1/2$, and therefore  
$Z_N(N/m)<2^{1-m}$ (the additional factor of 2 comes from RWs with $S_N<0$). 
Consequently, $Z_N(L)<2^{2-(N/L)}$ for any value of $N$ and $L$.
This establishes a strict upper bound, which is significant for 
small values of $\ell$:
\begin{equation}
z(\ell)<4\exp(-\ln 2/\ell).
\end{equation} 
It is possible to further improve on this upper bound,
and we will next demonstrate such an improvement which 
is by no means final.
Consider a pair of adjacent segments 
(e.g. $\omega_1$ and $\omega_2$) described above, with 
$S_L(\omega_1),S_L(\omega_2) >0$. 
Let $i$ be the {\it smallest} index where $S_i(\omega_1)=S_L(\omega_1)$,
and $j$ the {\it largest} index where $S_j(\omega_2)=0$.
In that case, the segment from $i$ to $L+j$ (on $\omega$) is a loop, 
and therefore $i>j$ since $\omega$ cannot have a loop larger than $L$.
For two randomly selected segments, this condition is satisfied
with probability $1/2$, which can be calculated from the
known probability 
distribution of ``last return to the origin''\cite{EKlongloop,mathnote}.
Since there are $m/2$ statistically independent adjacent pairs,
this observation further suppresses the upper bound on the probability
distribution by a factor of $2^{-m/2}$, improving the overall
upper bound to
\begin{equation}
z(\ell)<4\sqrt{2}\exp(-1.5\ln 2/\ell),
\end{equation} 
which makes the best (so far) analytical lower bound on the exponential 
factor $B>1.5\ln 2\approx 1.03972$.

In order to find a lower bound on the probability distribution,
let us again consider the sequence $\omega$ and its $m$ pieces
$\{\omega_i\}$ of length $L$ each. We'd like to construct each 
$\omega_i$ independently in such a way to guarantee that the resulting 
walk $\omega$ does not have loops larger than $L$. This can again
be done in many different ways, and the following is by no means 
optimal. The quality of the bound usually depends on how complicated
the specifications of each piece are, and the limiting factor 
seems to be the analytical tractability of the associated probabilities.
The following represents the best bound we have been able to establish
analytically.

\begin{figure}
\narrowtext
\centerline{\epsfxsize=2.9truein\epsffile{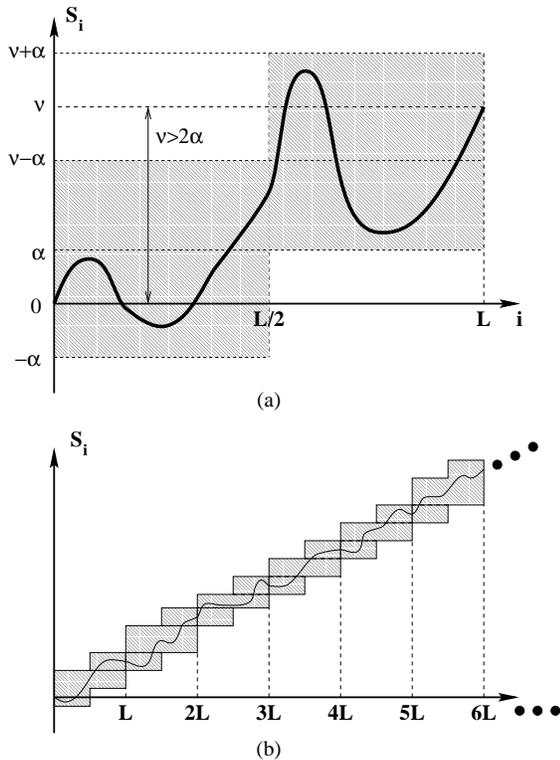}}
\caption{(a) Example of a walk that satisfies the conditions in 
Eq.(\protect\ref{eqlower}). Each such walk remains entirely 
within the shaded area. (b) When such walks are joined together,
the resulting walk does not have loops that are larger than
or equal to $L$, since such loops cannot fit in the shaded area.}
\label{lowerbound}
\medskip
\end{figure}

The specifications of each piece is as follows:
\begin{equation}
\label{eqlower}
\cases{-\alpha<S_i<S_L-\alpha,\;& $0<i\leq L/2$, \cr
       \alpha<S_i<S_L+\alpha,\;& $L/2< i\leq L$.}
\end{equation}
Figure~\ref{lowerbound}(a) shows these specifications graphically.
Clearly, $S_L>2\alpha$ is required.
Figure~\ref{lowerbound}(b) shows how the joining of such 
pieces results in a sequence $\omega$ that has no loops larger
than $L$.

The probability $p_s$ of meeting the stated specifications can be  evaluated 
numerically to high accuracy using the method of reflections\cite{mathnote} 
and summing over all possible values of $S_{L/2}$ and $S_L$ for a given 
$\alpha$. The largest value 
for the probability yields the tightest lower bound on $z(\ell)$,
so it is desirable to tune $\alpha$ in order to optimize the bound.
We pick $\alpha=0.5\sqrt{L}$, which is very close to the optimal
value. In that case, the probability for a RW to satisfy the
requirements (\ref{eqlower}) for large $L$ is $p_s \approx 0.031585$.
This yields
\begin{equation}
z(\ell)>2 p_s\exp(-\ln p_s/\ell)\approx0.06317 e^{-3.455/\ell}.
\end{equation}
Clearly, neither the upper nor the lower bounds we have established are
very tight, and they do not rule out the possibility of a power-law
prefactor. However, there is very
convincing numerical evidence that there is no power law prefactor
in $z(\ell)$, i.e. that $\lim_{\ell\to 0} z(\ell)=C\exp(-B/\ell)$ where
$C$ and $B$ are constants that are determined in the following
section.

\section{Numerical Work}
\label{numerical}

In this section, we present numerical studies to determine 
$p(\ell)$ and $z(\ell)$ in the $\ell \ll 1$ limit. As stated
earlier\cite{KEloop,EKlongloop}, a standard Monte Carlo method
of determining $p(\ell)$ from a random sample of all possible
walks is ineffective at probing $\ell\lesssim 0.15$, since the 
probabilities become very small. A similar problem arises 
when it is necessary to randomly sample very large self-avoiding
walks (SAWs) in two and three dimensions: 
The probability of generating a SAW is exponentially
small in its overall length, i.e. the probability of
picking a SAW out of RWs of length $N\ll 1$ scales as 
$P_{SAW}(N)\sim N^{\gamma}e^{-aN}$, where $a$ and $\gamma$ are
constants that depend only on the dimensionality of the SAW.
 A common way to circumvent this problem is to build large SAWs 
recursively by joining smaller SAWs. This method 
significantly reduces the number of operations needed by completely 
eliminating its dependence on the leading exponential factor:
The probability of creating a SAW of length $N$ by joining
two randomly selected SAWs of length $N/2$ scales only as
$N^{-\gamma}$, and the number of operations needed to generate
a randomly sampled SAW grows as $e^{\gamma(\log_2 N)^2/2}$
instead of $e^{aN}$. Of course, creating SAWs in one dimension is trivial,
but the extension of this method to one-dimensional walks is still very
useful for our problem, since creating RWs with very small loops
is similar to creating SAWs [in fact $P_{SAW}(N)=Z_N(1)$],
and can be used to sample $z(\ell)$ efficiently at small $\ell$. 

In this implementation of the algorithm,
we start from pairs of RWs of length $L$ (with nonzero total displacement)
and join them, keeping only resultant walks whose largest loops are 
smaller than $L$. At the first level, this creates walks that
contribute to $Z_{2L}(L)$, with equal probability. We then
iterate this process by pairing the resultant walks at each level.
After the $n$th level, we end up with a representative
sample of all walks that contribute to $Z_{2^nL}(L)$, which can then
be used to determine a histogram for the probability distributions
for $0<\ell<2^{-n}$.

\begin{figure}
\narrowtext
\centerline{\epsfxsize=2.9truein\epsffile{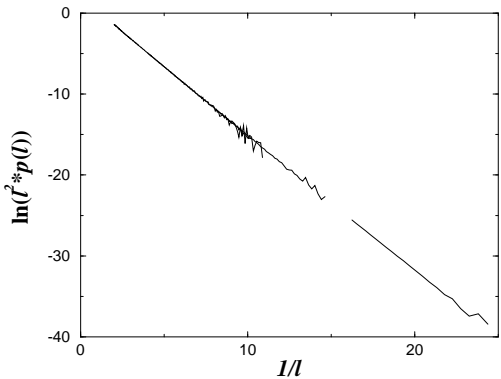}}
\caption{The probability density $p(\ell)$ for $0.04<\ell<1/2$ confirms the
suggested form (\protect\ref{eqpell}) down to probabilities as low as 
$10^{-15}$. The overall walk size is $N=2048$. 
Four (partially overlapping) plots were generated from 
runs that terminated after recursion levels $1$ through $4$.}
\label{psmallell}
\medskip
\end{figure}

We also need keep track of the probability of 
success $R_n$ at each level, which is given by
\begin{equation}
R_n(L)\equiv\frac{Z_{2^nL}(L)}{[Z_{2^{n-1}L}(L)]^2},
\end{equation} 
in order to determine the overall normalization of the probability 
distributions. One big advantage of studying one-dimensional walks
is that the probability of success $R_n(L)$ actually becomes 
{\it independent of} $n$, i. e., in the continuum limit
\begin{equation}
\label{recrel}
z(\ell)=R[z(2\ell)]^2, \; \ell\ll 1, 
\end{equation}
where $R=\lim_{L\to\infty}\lim_{n\to\infty}R_n(L)$ is a nonzero constant. 
(For the one-dimensional SAW, the probability of success is just $1/2$.) 
Typically, variations in $R_n(L)$ were 
within statistical fluctuations (0.1 to 0.3\%) for
$n\ge3$. 
When $R_n(L)$ is independent of $n$, the number of operations 
needed to sample a representative
walk that contributes to $z(\ell)$ is only polynomial in $\ell^{-1}$,
which speeds up the algorithm enormously.
Furthermore, this implies that for $\ell \ll 1$,
\begin{eqnarray}
z(\ell)&=&C\exp\{-B/\ell\}, \\
\label{eqpell}
p(\ell)&=&\frac{BC}{\ell^2}\exp\{-B/\ell\},
\end{eqnarray}
where $C=R^{-1}$ and $B$ are constants; there are no power-law
prefactors in $z(\ell)$.
This result can be verified numerically by looking at the results of 
the described recursive algorithm: Fig.~\ref{psmallell}
confirms the functional form (\ref{eqpell}) over about twelve decades
in the probability density $p(\ell)$, probing values of $\ell$ down to
0.04\cite{compfoot}. 
 
\begin{figure}
\narrowtext
\centerline{\epsfxsize=2.9truein\epsffile{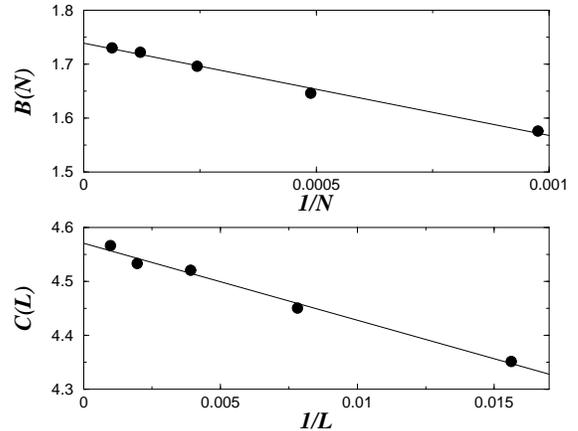}}
\caption{Size dependence of the constants that appear in the $L/N \ll 1$
limit of $Z_N(L)$ and $P_N(L)$. {\it Top:\ } The exponential constant 
$B(N)$ determined from plots of 
$\ell^2 p(\ell)$ as a function of total walk length $N$.
Statistical errors are smaller than symbol sizes.
{\it Bottom:\ }The prefactor $C(L)$ determined from success probabilities
$R_4(L)$ as a function of largest loop size $L$. Statistical errors 
are roughly the size of symbols.}
\label{finitesize}
\medskip
\end{figure}

The constants $C$ and $B$ in the continuum limit can be determined
accurately by plotting their dependence on walk length.
$C$ is simply the inverse of the success probability $R$ as mentioned
earlier, whereas $B$ is given by the slope of the graph in
Fig~\ref{psmallell}. Fig.~\ref{finitesize} shows these plots,
which yield
\begin{eqnarray}
C&=&4.57\pm0.01, \\
B&=&1.73\pm0.02.
\end{eqnarray}.

\section{related problems}
\label{related}

Behavior that is strikingly similar to those of $p(\ell)$ are 
exhibited by probability distributions of extremal properties 
in certain random systems. One simple example is a one--dimensional
random cutting process\cite{Derrida,Frach}: A unit interval
is cut at a randomly selected point (with uniform probability),
and the same cutting process is repeatedly applied to the interval that
remains to the {\it right} of the latest cut, ad infinitum. 
The probability distribution $p'(\ell)$ for the size of 
the {\it largest} interval that remains at the end of the 
cutting process exhibits singularities of the form 
$|\ell-1/k|^{k-1}$ at each value of $k$, which become 
progressively weaker as the integer $k$ is increased, 
leading to an essential singularity at $\ell=0$. 
The origin of these singularities 
can be traced to the fact that the pieces (among which the largest
one is chosen) constitute a {\it partition} of the entire interval,
which implies that the sum of the sizes of all pieces equals the 
size of the initial interval, which is 1. 
Consequently, any piece that is larger 
than $1/2$ is necessarily the largest, and
in general there can be at most $k-1$ pieces that are larger than
$1/k$. This causes singular behavior in $p'(\ell)$ at $\ell=1/k$
for all $k$. Similar ``sum rules'' apply to the all the other systems 
that are discussed in Ref.~\cite{Frach}.
However, this property is not satisfied by our problem, since
loops can and do overlap. 
We have numerically examined the vicinity of $\ell=1/3$, and
conclude that there are no singularities in the 
first and second derivatives of $p(\ell)$ with a prefactor
of $O(1)$. Although we cannot rule out the possibility of weaker
singularities or unusually small prefactors, the evidence seems
to suggest that they do not exist.

\section{Conclusion}

With the help of an efficient Monte Carlo algorithm and analytical
upper and lower bounds, we have clarified some of the issues surrounding 
the behavior of the probability density $p(\ell)$ for small values
of its argument, and we have been able to better understand
and characterize the essential singularity at $\ell=0$. 
In this limit,
the connection of this problem to SAWs becomes much more transparent,
and it is likely that this connection can be further exploited.

This work was supported by the National Science Foundation, 
by the MRSEC program through Grant DMR-9400396 and
through Grants DMR-9106237, DMR-9417047 and DMR-9416910;
and by the Israel Science Foundation founded by The
Israel Academy of Sciences under Grant No. 246/96.

\end{multicols}


\begin{references}
\bibitem{compfoot}In principle, it is straightforward to probe much
smaller values of $\ell$ since the algorithm is only polynomial
in $\ell^{-1}$. Our numerical results demonstrate what can be achieved
with runs that typically last a few days on a workstation such 
as SUN SPARCstation 5, and were sufficient for the purposes of this 
study.
\end{references}
\end{document}